\documentclass[prb,aps,showpacs,twocolumn,unsortedaddress,superscriptaddress]{revtex4-1}
\usepackage{graphics, bm, psfrag, amsmath, amssymb, epsfig, grffile, float, xspace}
\usepackage{subfigure}
\usepackage{color}

\newcommand{\mos}{${\rm MoS}_2$\xspace}
\newcommand{\dd}{{\bf d}}
\newcommand{\kk}{{\bf k}}
\newcommand{\hh}{{\bf h}}

\begin{document}

\title{General Conditions for Proximity-Induced Odd-Frequency Superconductivity in Two-Dimensional Electronic Systems}

\author{Christopher Triola}
\affiliation{Nordita, Center for Quantum Materials, KTH Royal Institute of Technology and Stockholm University, Roslagstullsbacken 23, 10691 Stockholm, Sweden}
\author{Driss M. Badiane}
\affiliation{Department of Physics, College of William and Mary, Williamsburg, Virginia 23187, USA}
\author{Alexander V. Balatsky}
\affiliation{Nordita, Center for Quantum Materials, KTH Royal Institute of Technology and Stockholm University, Roslagstullsbacken 23, 10691 Stockholm, Sweden}
\affiliation{Institute for Materials Science, Los Alamos National Laboratory, Los Alamos, New Mexico 87545, USA}
\author{E. Rossi}
\affiliation{Department of Physics, College of William and Mary, Williamsburg, Virginia 23187, USA}

\begin{abstract}

We obtain the general conditions for the emergence of odd-frequency superconducting pairing
in a two-dimensional (2D) electronic system proximity coupled to a superconductor, making minimal assumptions about both the 2D system and the superconductor. 
Using our general results we show that a simple heterostructure
formed by a monolayer of a group VI transition metal dichalcogenide, such as molybdenum disulfide,
and an $s$-wave superconductor with Rashba spin-orbit coupling exhibits odd-frequency
superconducting pairing. 
Our results allow the identification of a new class of systems among van der Waals heterostructures
in which odd-frequency superconductivity should be present.
%
%

\end{abstract}


\maketitle


Low-dimensional heterostructures hold the promise for new technologies\cite{xiang2006ge,miller2010graphene,zhu2011carbon,radisavljevic_natlett_2011,geim2013van} as well as granting us access to many unconventional quantum states including novel forms of superfluidity\cite{ZhangPRL2013}, manipulation of spin textures \cite{JinPRB2013,triola2014prl}, and unconventional superconductivity \cite{TriolaPRB2014,radisavljevic_natlett_2011,DemlerPRB1997,TokuyasuPRB1988,Black-SchafferPRB2013,LinderPRB2010_1,LinderPRL2009,parhizgar_2014_prb}.
In addition, theoretical analyses have shown that Majorana bound states
can appear in heterostructures incorporating superconducting materials 
%
\cite{fu2008superconducting,sau2010,lutchyn2010,oreg2010,hasan2010colloquium,qi2011rmp,alicea2013exotic,mourik2012signatures,rokhinson2012fractional,das2012zero,deng2012anomalous,finck2013anomalous,churchill2013superconductor}.
%
%
Given the variety of possible exotic states in low-dimensional heterostructures and that the fabrication of layered heterostuctures has rapidly advanced in recent years \cite{geim2013van} it is important to continue developing our understanding of their electronic properties. One important facet of this understanding is the classification of the possible symmetries of the proximity-induced superconductivity in these structures.       

The symmetries of a superconductor can be characterized by investigating the properties of the anomalous Green's function
$F_{\alpha\beta}(\textbf{r}_1,t_1;\textbf{r}_2,t_2)=\langle T c_\alpha(\textbf{r}_1,t_1)c_\beta(\textbf{r}_2,t_2)\rangle$, 
where $ c_\sigma(\textbf{r}_i,t_i)$ is the fermionic annihilation operator for an electron at position ${\bf r}_i$ time $t_i$
with spin $\sigma$, $T$ is the time ordering operator, and the angle brackets denote the expectation value.
%
%
%
Given the fermionic nature of the quasiparticles $F_{\alpha\beta}(\textbf{r}_1,t_1;\textbf{r}_2,t_2)=-F_{\beta\alpha}(\textbf{r}_2,t_2;\textbf{r}_1,t_1)$. 
Conventionally this is taken to imply that if the quasiparticle pair is in a spin singlet state then the pairing amplitude is even in parity while if it is a spin triplet the pairing amplitude is odd in parity. However, if the pairing amplitude is odd in time, or, equivalently, odd in frequency, 
spin triplet pairs can be even in parity and spin singlet pairs can be
odd in parity as was originally proposed for superfluid $^3\text{He}$ by Berezinskii\cite{Berezinskii1974} and later for superconductivity by Balatsky and Abrahams\cite{BalatskyPRB1992}. 

The study of odd-frequency SC has been hindered
by the scarcity of experimental systems in which it can be realized.
Soon after the original suggestion that in general an
odd-frequency pairing term could be present it was realized that
it would be challenging to get such a term via electron-phonon interactions
and that a spin-dependent electron-electron interaction would be necessary
\cite{abrahams1993}.
This fact greatly restricts the number of systems in which odd-frequency SC 
can be realized. 
However, in recent years it has become apparent that odd-frequency SC can be obtained in a variety of different types of heterostructures \cite{EschrigNat2008,LinderPRB2008,YokoyamaPRB2012,Black-SchafferPRB2012,Black-SchafferPRB2013,TriolaPRB2014,LinderPRL2009,LinderPRB2010_2,TanakaPRB2007,TanakaJPSJ2012,parhizgar_2014_prb,di2015signature}. The recent impressive explosion of realizable heterostructures has made the piecemeal approach unfeasible: a theoretical treatment able to provide the general conditions in which odd-frequency SC should be present in heterostructures has become necessary.
In this work we present such a general treatment.
Our general treatment also makes possible the identification of novel, somehow unexpected, engineered systems
in which such pairing should be present, as exemplified by the heterostructure formed by one monolayer of molybdenum disulfide (\mos) placed on superconducting Pb,
which we discuss in the second part of the Letter.
In particular, showing what are the necessary elements that a van der Waals heterostructure
must have to exhibit odd-frequency SC adds this important class of systems
%
to the odd-frequency playbook.
Our work also makes it possible to select among such systems the ones in which a
direct observation of the signatures due to odd-frequency SC is more readily achievable, for example, via scanning tunneling microscopy (STS) and angle resolved photoemission spectroscopy (ARPES).

The Hamiltonian ($H$) describing the most general heterostructure formed by a two-dimensional (2D) electron gas (2DEG) and a superconductor
can be written as  $H=H_{2D}+H_{SC}+H_t$ where
\begin{align}
H_{2D}&=\sum_{\textbf{k},\sigma,\sigma'}c^\dagger_{\textbf{k},\sigma}\left[h_0(\textbf{k})\sigma_0 + \textbf{h}(\textbf{k})\cdot\boldsymbol\sigma\right]_{\sigma,\sigma'}c_{\textbf{k},\sigma'} \label{eq:H2DEG} \\
H_{SC}&=\sum_{\textbf{k}\sigma\sigma'}d^\dagger_{\textbf{k}\sigma} h^{SC}_{\sigma\sigma'}(\textbf{k})d_{\textbf{k}\sigma'} 
+ \sum_{\textbf{k}\sigma\sigma'}d^\dagger_{\textbf{k}\sigma}\Delta_{\textbf{k}\sigma\sigma'}d^\dagger_{-\textbf{k}\sigma'} + \text{h.c.} 
 \label{eq:HSC}
\end{align} 
\begin{align}
H_t&=t\sum_{\textbf{k},\sigma} d^\dagger_{\textbf{k},\sigma}c_{\textbf{k},\sigma} + \text{h.c.}
\label{eq:Ht}
\end{align}
are the Hamiltonians describing the 2DEG, the superconductor, and the tunneling between the two systems, respectively. 
In Eqs.~(\ref{eq:H2DEG})-(\ref{eq:Ht}) 
$\sigma_0$ is the identity matrix in spin space, 
$\boldsymbol\sigma=(\sigma_1,\sigma_2,\sigma_3)$ is the vector of Pauli matrices in spin space, 
$c^{\dagger}_{\textbf{k},\sigma}$ ($d^{\dagger}_{\textbf{k},\sigma}$) and $c_{\textbf{k},\sigma}$ ($d_{\textbf{k},\sigma}$) are the creation and annihilation operators, respectively, acting on the fermionic states in the 2DEG (SC) layer with momentum $\textbf{k}$ and spin $\sigma$, 
$h_0(\textbf{k})$ is the spin-independent part of $H_{2D}$ and $\hh(\kk)$ is the field that describes
its spin-dependent part due to an exchange field and/or spin-orbit coupling,
%
$h^{SC}_{\sigma,\sigma'}(\textbf{k})$ describes the quasiparticle spectrum of the normal state of the superconductor,
$\Delta_{\textbf{k};\sigma,\sigma'}$ is the superconducting gap,
and $t$ is the tunneling between the 2D system and the SC. 
%
%
We assume that the tunneling conserves both spin and momentum given that this is the most common situation and that we wish to identify the most general condition to realize odd-frequency SC without having to resort to spin-active interfaces that are often difficult to realize experimentally.
%
%
%
To keep the treatment general we make no assumptions
on the form of $\hh(\kk)$, $h_{\sigma,\sigma'}^{SC}(\textbf{k})$, and $\Delta_{\textbf{k};\sigma,\sigma'}$. 
%

The anomalous Green's function associated with the superconductor described by Eq.~(\ref{eq:HSC}) is given by $\hat{F}^{SC}_{\textbf{k};i\omega_n}=\left[ \hat{\Delta}^\dagger_{-\textbf{k}} - \left(i\omega_n +\hat{h}^{SC}(-\textbf{k})^*\right)\hat{\Delta}^{-1}_{\textbf{k}}\left(i\omega_n -\hat{h}^{SC}(\textbf{k})\right) \right]^{-1}$. We can parametrize this matrix in terms of singlet and triplet parts,
\begin{equation}
\hat{F}^{SC}_{\textbf{k};i\omega_n}=\left(s^{SC}_{\textbf{k},i\omega_n}\sigma_0 + \textbf{d}_{\textbf{k},i\omega_n}\cdot\boldsymbol\sigma \right)i\sigma_2
\label{eq:FSC}
\end{equation}
where  $\omega_n$ is the Matsubara frequency, and
$s^{SC}_{\textbf{k},i\omega_n}$ and the three-component complex vector $\textbf{d}_{\textbf{k},i\omega_n}$~\cite{mackenzie2003} give the singlet and triplet superconducting amplitudes, respectively. 
The leading order contributions to the proximity-induced superconducting pairing in the 2DEG are given by:
\begin{equation}
\hat{F}^{2D}_{\textbf{k};i\omega_n} =t^2 \ \hat{G}^{2D}_{\textbf{k};i\omega_n} \ \hat{F}^{SC}_{\textbf{k};i\omega_n} \ \left(\hat{G}^{2D}_{-\textbf{k};-i\omega_n}\right)^T
\label{eq:F2DEG}
\end{equation}
where 
\begin{equation}
\hat{G}^{2D}_{\textbf{k};i\omega_n}=\dfrac{(i\omega_n-h_0(\textbf{k}))\sigma_0+\textbf{h}(\textbf{k})\cdot\boldsymbol\sigma}{(i\omega_n-h_0(\textbf{k}))^2 - |\textbf{h}(\textbf{k})|^2}
\end{equation}
is the Green's function associated with the 2DEG.

It is convenient to separate the anomalous Green's function $\hat{F}^{2D}_{\textbf{k};i\omega_n}$ 
into two parts 
$\hat{F}^{2D}_{\textbf{k};i\omega_n} = A_{\textbf{k};i\omega_n} \left(F_{\textbf{k};i\omega_n}^{odd} + F_{\textbf{k};i\omega_n}^{even} \right)$ 
where 
$A_{\textbf{k};i\omega_n}$
is generally a function even in $\omega_n$~\cite{sm}, and
$F_{\textbf{k};i\omega_n}^{odd}$ and $F_{\textbf{k};i\omega_n}^{even}$ are the odd- and even-frequency $2\times 2$ matrices 
describing the spin structure of the induced superconducting pairs respectively.

Let $\textbf{h}_{\pm}(\textbf{k})\equiv \textbf{h}(\textbf{k})\pm \textbf{h}(-\textbf{k})$.
Then for $F_{\textbf{k};i\omega_n}^{even}$ we find
%
\begin{equation*}
%
%
F_{\textbf{k};i\omega_n}^{even} = \left(S_{\textbf{k};i\omega_n}^{even}\sigma_0 + {\bf D}_{\textbf{k};i\omega_n}^{even}\cdot\boldsymbol\sigma \right)i\sigma_2 
\end{equation*}
where $S_{\textbf{k};i\omega_n}^{even}$, ${\bf D}_{\textbf{k};i\omega_n}^{even}$ are the singlet and triplet components, respectively, given by:
\begin{equation}
\begin{aligned}
S_{\textbf{k};i\omega_n}^{even} = &\left[ \omega_n^2 +h_0^2(\textbf{k}) 
                            -\frac{1}{4}(|\hh_+(\kk)|^2 - |\hh_-(\kk)|^2) \right]s_{\textbf{k};i\omega_n}^{SC} \\
                          -&\left[h_0(\kk)\hh_-(\kk)+\frac{i}{2}\hh_+(\kk)\times\hh_-(\kk)\right]\cdot \textbf{d}_{\textbf{k};i\omega_n}\\
{\bf D}_{\textbf{k};i\omega}^{even}= &\left[ \omega_n^2 + h_0^2(\textbf{k}) + \frac{1}{4}(|\hh_+(\kk)|^2 - |\hh_-(\kk)|^2)\right]\textbf{d}_{\textbf{k};i\omega_n} \\
                          -&ih_0(\textbf{k})\hh_+(\kk)\times\textbf{d}_{\textbf{k};i\omega_n} -\frac{1}{2}\hh_+(\kk)\left(\hh_+(\kk)\cdot\textbf{d}_{\textbf{k};i\omega_n}\right) \\
                                +&\frac{1}{2}\hh_-(\kk)\left(\hh_-(\kk)\cdot\textbf{d}_{\textbf{k};i\omega_n}\right) \\
                               -&\left[h_0(\kk)\hh_-(\kk)-\frac{i}{2}\hh_+(\kk)\times\hh_-(\kk)\right]s_{\textbf{k};i\omega_n}^{SC}. 
%
%
\end{aligned}
\label{eq:Phi_e}
\end{equation}
%
The first line (three lines) of the expression for $S_{\textbf{k};i\omega_n}^{even}$ (${\bf D}_{\textbf{k};i\omega_n}^{even}$)
shows that, as expected, a singlet (triplet) pairing is induced, via the proximity effect, in the 2DEG by 
a singlet (triplet) superconductor, regardless of the value of $\hh$. 
The last line for the expression of $S_{\textbf{k};i\omega_n}^{even}$ (${\bf D}_{\textbf{k};i\omega_n}^{even}$)
shows that if $\hh_-\neq 0$, by the proximity effect, then in the 2DEG we have even-frequency superconductivity
with both singlet and triplet pairing even if the substrate superconductor only has singlet or triplet pairing. 
It also shows that the strength of the pairing in the 2DEG with spin structure different from that of the substrate
is proportional to $\hh_-(\kk)$ and is augmented when $\hh_-\times\hh_+\neq 0$.
This result shows how the presence of spin-orbit coupling, which gives rise to $\hh_-\neq 0$,
qualitatively affects the nature of the conventional (even-frequency) superconducting pairing
induced by proximity.
We then find that the interplay of the field $\hh$ {\em in the 2DEG} and the superconducting pairing
in the substrate gives rise to an odd-frequency pairing term,
%
%
%
\begin{equation*}
F_{\textbf{k};i\omega_n}^{odd} = i\omega_n\left(S_{\textbf{k};i\omega_n}^{odd}\sigma_0 + {\bf D}_{\textbf{k};i\omega_n}^{odd}\cdot\boldsymbol\sigma \right)i\sigma_2 
\end{equation*}
with $S_{\textbf{k};i\omega_n}^{odd}$, ${\bf D}_{\textbf{k};i\omega_n}^{odd}$ being the odd-frequency singlet and triplet components, respectively, given by:
\begin{equation}
\begin{aligned}
S_{\textbf{k};i\omega_n}^{odd} &=- \textbf{h}_{+}(\textbf{k})\cdot\textbf{d}_{\textbf{k};i\omega_n} \\
{\bf D}_{\textbf{k};i\omega}^{odd} &=- \textbf{h}_{+}(\textbf{k})s_{\textbf{k};i\omega_n}^{SC} - i\textbf{h}_{-}(\textbf{k})\times\textbf{d}_{\textbf{k};i\omega_n}.
\end{aligned}
\label{eq:Phi_o}
\end{equation}
%
This result clearly shows that it {\em is} possible to get an odd-frequency singlet term provided
the substrate is a triplet superconductor with a $\dd$ vector that is not perpendicular to the even component 
of $\hh$, $\hh_+$. 
Notice that because $\hh$ and $\dd$ belong to different layers they are not constrained to be in
any specific relation.
Equation~(\ref{eq:Phi_o}) also shows that an odd-frequency triplet term is present if
both $\hh_+$ and the singlet pairing in the substrate $s^{SC}$ are not 0, 
as shown previously~\cite{EschrigNat2008,LinderPRB2008,TriolaPRB2014}.
Equation~(\ref{eq:Phi_o}) therefore shows that when $\hh_+\neq 0$, and $\hh_-=0$, by the proximity effect,
we have odd-frequency superconductivity in the 2DEG that has the ``opposite'' spin structure
from the superconductivity in the substrate: triplet if the substrate is a singlet superconductor,
singlet if the substrate is a triplet superconductor (with $\dd$ not orthogonal to $\hh_+$).
A very interesting and novel result is that even when $\hh_+=0$, i.e. no ferromagnetism is present in the 2DEG,
we can have odd-frequency superconductivity in the 2DEG, {\em without having to assume the presence of a spin-active interface},
if the 2DEG has spin-orbit coupling, so that $\hh_-\neq 0$, and the substrate is a triplet superconductor with $\dd$
not parallel to $\hh_-$ (again, we emphasize that because 
$\hh_-$ and $\dd$ belong to different layers they are not locked to each other).
This is a result that significantly enlarges the set of engineered structures in which to
look for odd-frequency superconductivity by adding a whole new class of heterostructures.
As we show below, a system that falls into this class is a heterostructure formed by 
a group-VI dichalcogenide monolayer and a superconductor's surface with Rashba spin-orbit coupling.

\begin{figure}
 \begin{center}
  \centering
  \includegraphics[width=8.0cm]{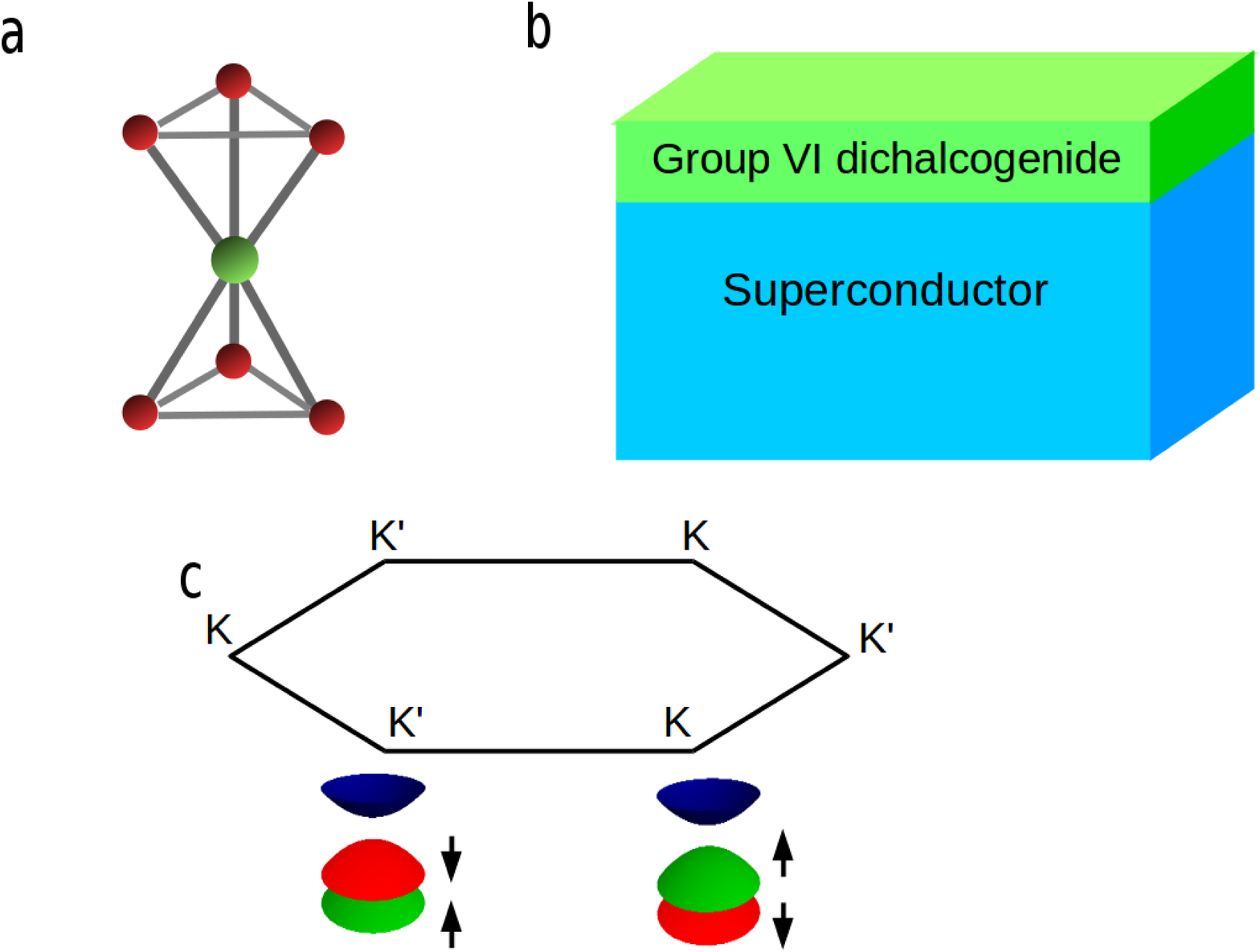}
  \caption{
           (Color online) a) Unit cell for a monolayer TMD. A single monolayer is composed of three covalently bonded layers trigonally coordinated with a layer of transition metal sandwiched between two layers of chalcogen. b) Schematic of a heterostructure formed by exfoliating a TMD monolayer onto a superconductor. c) Sketch of the band structure of a TMD monolayer with the $d$-electron bands appearing at the $K$ and $K'$ points with a band gap of 1.8eV separating a pair of spin-degenerate conduction bands from a pair of spin-polarized bands. Notice that the polarization is different in the two inequivalent valleys ($K$ and $K'$).  
         }
  \label{fig:figure1}
 \end{center}
\end{figure}

Transition metal dichalcogenides (TMDs), such as MoS$_2$,
have recently received a lot of attention due to their unusual electronic properties
and potential for applications in electronics.
\mos can be exfoliated down to monolayer 2D crystals\cite{novoselov_pnas_2005,mak_prl_2010,radisavljevic_natlett_2011,ross2013electrical}. These monolayers have been shown to possess a direct band gap of 1.8eV \cite{radisavljevic_natlett_2011,splendiani_nanolett_2010}, they can be gated \cite{radisavljevic_natlett_2011}, and have exhibited electron mobilities as high as 200 cm$^2$V$^{-1}$s$^{-1}$\cite{radisavljevic_natlett_2011}.  
Furthermore, the $d$-electron states exhibit a valley degree of freedom that is coupled to the electron spin\cite{XiaoPRL2012,mai2013many,xu2014spin}. 
In the context of our problem, this material is of great interest not only because it is a two-dimensional material that is readily available, easily manufactured and incorporated into heterostructures, but also because of its strong spin-orbit coupling.

Consider the heterostructure shown in Fig.~\ref{fig:figure1} composed of a TMD monolayer on top of a superconductor. 
The low-energy electronic states of a TMD monolayer are well described by the following valley-dependent Hamiltonian \cite{XiaoPRL2012}:
\begin{equation}
\begin{aligned}
\hat{H}^{TMD}_{\textbf{k},\lambda}&=\left[a\gamma\left( \lambda k_x \tau_1 + k_y \tau_2 \right) + \dfrac{u}{2} \tau_3 - \mu \tau_0\right]\otimes \sigma_0 \\
&-\dfrac{\lambda\alpha}{2}\left( \tau_3 - \tau_0 \right)\otimes \sigma_3
\end{aligned}
\label{eq:HDC}
\end{equation}
where $\tau_i$ are Pauli matrices acting on the orbital space of the TMD monolayer, $a$ is the lattice constant, $\gamma$ is the effective hopping integral, $u$ is the energy gap between the valence and conduction bands, $\alpha$ is the strength of the spin-orbit coupling, 
$\lambda=\pm 1$ is the valley index ($\lambda=1$ denotes the $K$ valley and $\lambda=-1$ denotes the $K'$ valley; see Fig.~\ref{fig:figure1}), 
$\textbf{k}=(k_x,k_y,0)$ is a vector describing small deviations from the $K$ or $K'$ point in $k$ space, 
and $\mu$ is the chemical potential. For MoS$_2$, $a=3.193\text{ \AA}$, $\gamma=1.10\text{ eV}$, $u=1.66\text{ eV}$, and $2\alpha=0.15\text{ eV}$ \cite{XiaoPRL2012}.

The Hamiltonian in Eq.~(\ref{eq:HDC}) possesses four eigenstates at the K and K' points, as shown in Fig.~\ref{fig:figure1}: two spin-degenerate conduction states separated by an eV-scale gap from two spin-polarized valence states.
For our analysis the most interesting case is when MoS$_2$ is hole doped.
For this reason in the following we use an effective two-band model in which we include only the valence bands. 
Considering the large gap between the valence and conduction bands this does not introduce any inaccuracy.
For small $k$ the valence band Hamiltonian can be written in spin space as
\begin{equation}
\hat{H}^{TMD}_{\textbf{k},\lambda}=-\left( \dfrac{a^2\gamma^2}{u}k^2 + \dfrac{u}{2} + \mu\right) \sigma_0 + \lambda\alpha\sigma_3.
\label{eq:h_dc}
\end{equation}

Notice that, taking into account the valley index $\lambda$, for the 
parity operator, $\mathcal{P}$, 
acting on a function, $f(\textbf{k},\lambda)$
we have $\mathcal{P}f(\textbf{k},\lambda)=f(-\textbf{k},-\lambda)$.
Using the notation used in Eqs~(\ref{eq:Phi_e}) and~(\ref{eq:Phi_o}) we then find that  
in this case 
$h_0(\textbf{k})=-\left(\dfrac{a^2\gamma^2}{u}k^2 + \dfrac{u}{2} + \mu\right)$,
$\textbf{h}_{+}(\textbf{k},\lambda)=0$, 
and  
$\textbf{h}_{-}(\textbf{k},\lambda)=2\lambda\alpha\hat{z}$, 
where $\hat{z}$ is the unit vector normal to the TMD monolayer.
%
Starting from the general Eqs~(\ref{eq:Phi_e}) and~(\ref{eq:Phi_o}) we then find
\begin{equation}
\begin{aligned}
S_{\textbf{k},\lambda;i\omega_n}^{even} &= \left( \omega_n^2 + \xi_{\textbf{k}}^2 +\alpha^2 \right)s_{\textbf{k},\lambda;i\omega_n}^{SC}-2\lambda\alpha\xi_{\textbf{k}}\hat{z}\cdot \textbf{d}_{\textbf{k},\lambda;i\omega_n} \\
{\bf D}_{\textbf{d},\lambda;i\omega}^{even} &=\left( \omega_n^2 + \xi_{\textbf{k}}^2 -\alpha^2 \right)\textbf{d}_{\textbf{k},\lambda;i\omega_n} +2\alpha^2\left(\hat{z}\cdot\textbf{d}_{\textbf{k},\lambda;i\omega_n}\right)\hat{z} \\
&- 2\lambda\alpha\xi_{\textbf{k}}s_{\textbf{k},\lambda;i\omega_n}^{SC} \hat{z}
\end{aligned}
\label{eq:Phi_e_dc}
\end{equation}
and
\begin{equation}
\begin{aligned}
S_{\textbf{k},\lambda;i\omega_n}^{odd} &=0 \\
{\bf D}_{\textbf{k},\lambda;i\omega}^{odd} &= -i2\lambda\alpha \hat{z}\times\textbf{d}_{\textbf{k},\lambda;i\omega_n}
\end{aligned}
\label{eq:Phi_o_dc}
\end{equation}
In accordance with Eq.~(\ref{eq:Phi_o}) we find that, given that $\hh_+=0$, to get odd-frequency superconductivity in the TMD we need
a substrate with nonzero triplet superconducting pairing.
In general this situation is realized in noncentrosymmetric 
superconductors. Additionally, this condition can be realized at the surface of centrosymmetric singlet superconductors with spin-orbit coupling since the surface breaks inversion
symmetry leading to the appearance of a Rashba spin-orbit term
that in turn induces a superconducting triplet component\cite{GorkovPRL2001}. This is expected to be the case for the surface of superconducting Pb.

Considering the case in which the superconductor in Fig.~\ref{fig:figure1}~(b) has Rashba spin-orbit coupling, the Hamiltonian matrix describing the single particle spectrum of the superconductor is $\hat{h}^{SC}(\textbf{k})=\epsilon_{\overline{\textbf{k}}}\hat{\sigma}_0+\eta\hat{z}\cdot(\boldsymbol\sigma\times\overline{\textbf{k}})$
where $\eta$ is the Rashba spin-orbit coupling in the superconductor surface, $\epsilon_{\overline{\textbf{k}}}$ is the dispersion of the normal state quasiparticles in the absence of spin-orbit coupling, and $\overline{\textbf{k}}$ is the momentum measured from the Brillouin zone center.
Considering that the dominant pairing is intraband we obtain~\cite{sm,GorkovPRL2001}
%
%
\begin{equation}
\hat{F}^{SC}_{\overline{\textbf{k}};i\omega_n}= \frac{\Delta}{(s_{\overline{\textbf{k}};i\omega_n}^{SC})^2-|{\bf d}_{\overline{\textbf{k}}}|^2} 
(s_{\overline{\textbf{k}};i\omega_n}^{SC}\sigma_0+{\bf d}_{\overline{\textbf{k}}}\cdot\boldsymbol\sigma)i\sigma_2
\label{eq:FSC_rashba}
\end{equation}
where $\Delta$ is the substrate's superconducting gap,
$s_{\overline{\textbf{k}};i\omega_n}^{SC}= \Delta^2 +\omega_n^2 + \epsilon_{\overline{\textbf{k}}}^2 + \eta^2\overline{k}^2$ and $\dd_{\overline{\textbf{k}}}=2\epsilon_{\overline{\textbf{k}}}\eta(-\overline{k}_y,\overline{k}_x,0)$. 
The key point of Eq.~(\ref{eq:FSC_rashba}) is that thanks to the Rashba spin-orbit coupling induced by the 
breaking of the inversion symmetry at the surface of the Pb substrate a triplet term appears in the $\hat F^{SC}$
and in addition the $\dd$ vector for such a triplet component is perpendicular to the field $\hh_-$ in the TMD monolayer.
The interplay of such a triplet component with the spin-orbit coupling {\em of the TMD monolayer} 
gives rise to odd-frequency SC in the TMD.

With the above definitions we can follow the same steps leading to Eqs~(\ref{eq:Phi_o}) and ~(\ref{eq:Phi_e}) and obtain the leading order contribution to the proximity-induced anomalous Green's function in the TMD layer as 
$
\hat{F}^{TMD}_{\textbf{k},\lambda;i\omega_n} = A^{TMD}_{\textbf{k},\lambda;i\omega_n} \left(F_{\textbf{k},\lambda;i\omega_n}^{odd} + F_{\textbf{k},\lambda;i\omega_n}^{even} \right)
$
where 
$$A^{TMD}_{\textbf{k},\lambda;i\omega_n}=\frac{\Delta t^2}{[(i\omega_n-\xi_\textbf{k})^2-\alpha^2]^2[(s_{\textbf{k}+\textbf{K}_\lambda;i\omega_n}^{SC})^2-|\dd_{\textbf{k}+\textbf{K}_\lambda}|^2]}.
$$
For the even-frequency singlet and triplet components of $\hat F^{TMD}$ we find
\begin{equation}
\begin{aligned}
S_{\textbf{k},\lambda;i\omega_n}^{even} &= \left( \omega_n^2 + \xi_{\textbf{k}}^2 +\alpha^2 \right)s_{\textbf{k}+\textbf{K}_\lambda;i\omega_n}^{SC} \\
{\bf D}_{\textbf{k},\lambda;i\omega}^{even} &=-\left( \omega_n^2 + \xi_{\textbf{k}}^2 -\alpha^2 \right)\dd_{\textbf{k}+\textbf{K}_\lambda} - 2\lambda\alpha\xi_{\textbf{k}}s_{\textbf{k}+\textbf{K}_\lambda;i\omega_n}^{SC} \hat{z}
\end{aligned}
\label{eq:Phi_e_dc_final}
\end{equation}
%
Given that $\hh_+=0$ [see Eq.~(\ref{eq:Phi_o_dc})], the odd-frequency singlet component vanishes whereas for the triplet
component we find
\begin{equation}
\begin{aligned}
{\bf D}_{\textbf{k},\lambda;i\omega}^{odd} &= i4\lambda\alpha\eta \epsilon_{\textbf{k}+\textbf{K}_\lambda} (\textbf{k}+\textbf{K}_\lambda)
\end{aligned}
\label{eq:Phi_o_dc_final}
\end{equation}
where $\textbf{K}_\lambda$ is the momentum vector at the $K$ ($K'$) point for $\lambda=1$ ($\lambda=-1$).  
Equation~(\ref{eq:Phi_o_dc_final}) shows that in the TMD the odd-frequency triplet component has a $d$-vector
pointing in the direction of the momentum. One can verify that this corresponds to an equal-spin spin triplet amplitude given by $F^{TMD}_{\uparrow\uparrow/\downarrow\downarrow}\sim i\omega_n\eta\alpha\epsilon_{\overline{\textbf{k}}}\lambda\left(\overline{k}_y\pm i\overline{k}_x\right)$ which is proportional to the product of the spin-orbit couplings in the two materials. 
%
%
Consistent with the general case,
we see that the emergence of this term requires the spin-orbit couplings in the two media to be nonparallel.

Our results  add a new class of systems, van der Waals (VDW) heterostructures, to the
odd-frequency playbook. van der Waals systems have many advantages:
(i)   the 2DEG in which odd-frequency pairing is present lives in a layer
     with an exposed surface, a fact that allows for ideal STS and ARPES
     measurements;
(ii)  as shown by the example of the \mos/Pb heterostructure, 
     it is possible to realize VDW systems with no ferromagnetic layers, or spin-active interfaces
     that exhibit odd-frequency SC;
(iii) the 2DEG in which odd-frequency pairing is present can be just one
     atom thick;
     this fact removes many of the complications associated with the
     interpretation of STS and ARPES data done in heterostructures
     in which each layer is several nanometers thick;
(iv) because the top layer is just one atom thick the electrons are
     truly confined in 2D; this fact, combined with the fact that
     according to our results the top layer can be a semiconductor,
     rather than a ferromagnetic metal as in previous proposals,
     ensures that the density of states (DOS) of the normal state is quite low and therefore
     allows for an easier observation of the features in the DOS due
     to the presence of odd-frequency pairing.

In conclusion, in this work we investigated the symmetries of proximity-induced superconducting pairing amplitudes in a 2DEG coupled to a superconductor. We arrived at a general expression relating the induced pairing amplitudes to the components of the anomalous Green's function of the superconducting substrate and the elements of the 2DEG Hamiltonian matrix,  $\hat{h}(\textbf{k})=h_0(\textbf{k})\hat{\sigma}_0+\textbf{h}(\textbf{k})\cdot\boldsymbol\sigma$. 
We have shown that the interplay of the spin-orbit coupling in the 2DEG and the superconducting pairing of the substrate
can give rise, via the proximity effect, to unusual superconducting pairings in the 2DEG.
We find that even when no ferromagnetism is present in the 2DEG, and there is no spin-active interface,
odd-frequency superconductivity can be induced in the 2DEG provided the 2DEG has spin-orbit coupling 
and the substrate has some triplet superconductivity. We then showed that this condition can be realized
in a \mos/Pb heterostructure. This result, combined with the general equations that we obtain,
adds a new class of systems, VDW heterostructures, to the
odd-frequency playbook.


\textit{Acknowledgements}. We thank Annica Black-Schaffer, Matthias Eschrig, Satrio Gani, Martin Rodriguez-Vega, Yudistira Virgus, and Junhua Zhang for useful discussions. The work of C.T. and A.V.B. was supported by the European Research Council (ERC) DM-321031 and US DOE BES E304, D.M.B. and E.R. acknowledge support from ONR and NSF CAREER Grant DMR-1455233.

\bibliographystyle{apsrev}
\bibliography{Dichal_Proxy}

\end{document}